\documentclass{elsart}

\usepackage{graphicx}

\usepackage{amssymb}

\begin{document}

\begin{frontmatter}



\title{Risk and Utility in Portfolio Optimization}


\author[label1]{Morrel H. Cohen}
\author[label2]{and Vincent D. Natoli}
\address[label1]{Dept. of Physics and Astronomy, Rutgers University, Piscataway, NJ 08854-8019, USA}

\address[label2]{ExxonMobil Research and Engineering, Route 22 East Annandale NJ 08886}

\begin{abstract}
Modern portfolio theory(MPT) addresses the problem of determining the optimum allocation of investment resources among a set of candidate assets.  In the original mean-variance approach of Markowitz, volatility is taken
as a proxy for risk, conflating uncertainty with risk. There have been many subsequent attempts to alleviate that weakness which, typically, combine utility and risk.  We present here a modification of MPT based on the inclusion of separate risk and utility criteria. We define risk as the probability of failure to meet a pre-established investment goal.  We define utility as the expectation of a utility function with positive and decreasing marginal value as a function of yield.  The emphasis throughout is on long investment horizons for which risk-free assets do not exist.  Analytic results are presented for a Gaussian probability distribution.  Risk-utility relations are explored via empirical stock-price data, and an illustrative portfolio is optimized using the empirical data. 

\end{abstract}

\end{frontmatter}

\section{Introduction}
\label{}

Two of the main pillars of mathematical finance are modern portfolio theory 
(MPT) and the Capital Asset Pricing Model(CAPM). The seminal work on MPT is attributed to
Markowitz who presented his mean-variance approach to asset allocation in 1952\cite{markovitz52}. It was soon amplified by 
Sharpe in 1964\cite{sharpe64} and by Lintner in 1965\cite{lintner65} with the introduction of the concept of the capital
market line and subsequent development of the CAPM. MPT permeates the teaching and practice of classical financial theory. Substantial 
portions of most textbooks on finance are devoted to it and its implications. Its influence 
has been profound.

The notion that portfolio volatility, the square root of the variance of the portfolio 
yield, is an adequate proxy for risk is fundamental to MPT. Similarly, the notion that 
there exists at least one risk - free asset is fundamental to the construction of the capital 
market line and the formulation of the CAPM. In the present paper, we discuss issues surrounding both of these notions and, abandoning 
them, introduce a novel method of portfolio optimization. The notion that variance 
measures risk is now viewed as a weak compromise with economic reality. Variance 
measures uncertainty, and there are circumstances of interest in which great uncertainty 
implies little risk. Similarly, supposing that there are risk-free assets or, more precisely, 
assets with unvarying yield is a poor approximation, particularly for long-time horizons.

There have been attempts to develop MPT with alternative definitions of risk, including a semi-variance, RMS loss, average downside risk, value at risk (VAR) and others\cite{markovitz59,var1,var2,var3} but to our knowledge, none is based on the classic notion that the probability of failure to meet a 
preset goal is the proper quantitative measure of risk
or on the elimination of the notion of a risk-free asset. 

In the following sections we give a brief introduction of MPT
with critiques of each of the above two fundamental notions. 
We show that the probability of success can be interpreted as an expected utility that is
deficient in some desirable features.  We construct an additional utility with the desired properties
and include it in the portfolio optimization.
We discuss how to define a real portfolio optimization 
problem using historical data and report the result of our risk and utility 
evaluation using the daily closing prices for 13,000 stocks listed on the NYSE and 
NASDAQ during the period 1977-1996. 
We conclude by presenting the results of our optimization for a 
portfolio drawn from a subset of low-risk, high utility stocks and
discuss the implications of our main findings.

\section{Modern Portfolio Theory}

The asset allocation problem is one of the fundamental concerns of financial 
theory. It can be phrased as a question: What is the optimal allocation of funds $F$ among a 
set of assets $\{A_i\}$ for a predetermined level of risk? Phrased in this way it leaves 
unspecified the meaning of optimality and of risk. Modern portfolio theory offers a two 
step answer via a particular specification of optimality and risk.            
The first step was taken in 1952 with the introduction of the mean-variance approach of Markowitz\cite{markovitz52}. By equating risk with variance, Markowitz derived an efficient frontier of portfolios which maximize return for given risk and opened the door to further advances in this theoretical framework.

The addition of a risk-free asset by Sharpe\cite{sharpe64} and Lintner\cite{lintner65} in the mid 1960s led to the capital market line and the CAPM. They supposed that 
there exists a risk-free asset $A_0$, whose yield, $Y_0$, did not fluctuate.
The line drawn from $Y_0$ tangent to the efficient 
frontier is then the locus of yields of all optimal portfolios which can be constructed by 
adding the risk-free asset to holdings drawn from the tangent set which via equilibrium arguments
is indentified with the market portfolio.  Along that line, termed the capital market line, return increases linearly
with risk\cite{luenberger}. The one-fund theorem follows, stating that any portfolio on the 
capital market line may be constructed from a combination of the risk-free asset and the market
portfolio\cite{tobin58}. 

\section{A Critique}

We have two fundamental objections to MPT.  First, volatility measures the uncertainty 
of yield.  While it may be positively correlated with risk in some cases, it does not, in 
general, measure risk.  Suppose, for example, that the specific goal for the portfolio is 
that its mean yield $\bar{Y}$ must equal or exceed a minimum acceptable value $Y_M$.  Suppose 
also that the volatility of the portfolio is large, perhaps significantly larger than $Y_M$.  
Nevertheless, if the mean yield is also large, enough so that $\bar{Y}-Y_M$ significantly 
exceeds the volatility, $V$, the probability that the goal is not met will be small.  There can be large 
uncertainty with little risk.  

Second, no asset is risk-free in the long term, and, depending on risk tolerance, perhaps 
not even in the short term.  There are various risks associated with any supposedly 
risk-free asset, including e.g., inflation risk, interest-rate risk, and exchange-rate risk.  We 
conclude that $A_0$, the candidate risk-free asset, should be added to the asset mix and optimized with
the rest.  The results of this addition are far reaching.  The efficient frontier is modified, the 
capital market line disappears, and the one-fund theorem is in general not valid.
The efficient market portfolio is no longer unique.

\section{Optimizing Via The Probability Of Success}
\subsection{Risk As The Probability Of Failure}
For more than four centuries, the \it probability of failure \rm has been taken as the quantitative 
measure of risk\cite{bernstein}.  Let $p$ be the probability of success. Then the risk is $1-p$ and the adverse 
odds $(1-p)/p$.  However, to define success there must be a goal.  We take as the goal of 
asset allocation the one previously introduced, namely that the expected portfolio yield 
equal or exceed $Y_M$, a minimum acceptable yield.  

The average yields and volatilities of the individual assets depend on $T_H$, the investment horizon 
(holding period) as of course does the entire probability distribution of $Y_i$, $P(Y_i)$.  We 
have investigated the dependence of $P(Y_i)$ and the volatility, $V_i$, on $T_H$ for 13,000 
stocks using price-time data from 1977-1996.  We found a non-universal power law dependence of 
$V_i$ on $T_H$, $V_i \propto {T_H}^{-s_i},\;0.5\leq s_i\leq 1.0$
We conclude that specifying $T_H$ is an essential part of defining the goal.  The minimum acceptable, yield $Y_M$, has two components, which must be specified 
independently, $Y_M=Y_M^0+\Delta Y_M$ where, $Y^0_M$ is the minimum acceptable real (deflated) after-tax yield.  $\Delta Y_M$ is an 
allowance for transaction costs, inflation, and tax costs.  
\subsection{Risk Tolerance And Portfolio Optimization}
In MPT, selecting a value for the volatility, $V$, of the optimal portfolio establishes 
uncertainty tolerance.  Instead, we specify a minimum acceptable value of $p$, $p^*$.  Stating 
that $p$ must equal or exceed $p^*$, establishes our risk tolerance.  With $\{A_i\}$, $Y_M$, and $T_H$ 
chosen and with knowledge of $P[\{Y_i\}\vert T_H]$ the joint probability distribution of the $Y_i$, $p$ can be 
evaluated for each allocation vector, $\{X_i\}$, 
\begin{equation}
\label{eqn3}
p=\int^\infty_{Y_M}dY\;P[Y\vert T_H].
\end{equation}
Optimization then consists of finding the supremum of $\bar{Y}$ for $p\geq p^*$ subject to 
$\sum_i X_i=1,\;\sum_i X_i \bar{Y}_i=\bar{Y}$.
\section{Risk And Utility}
\subsection{The Probability of Success as a Utility}
The definition of $p$ given by Eq. (\ref{eqn3}) can be rewritten as 
\begin{equation}
\label{eqn4}
p=\int dY\: U_p(Y-Y_M)P(Y)=\langle U_p(Y-Y_M)\rangle,
\end{equation}
that is as the expectation value of the Heaviside unit function,
\begin{equation}
\label{eqn5}
U_p(\zeta)=1,\hspace{.1in}\zeta\geq 0;\;\;\; U_p(\zeta)=0,\hspace{.1in}\zeta < 0.
\end{equation}
In this form, it can be interpreted as an expected \it utility \rm with $U_p(Y-Y_M)$ the \it utility 
function\rm. $U_p(Y-Y_M)$ punishes all losses equally and thus lacks the ability to discriminate.
Thus $p$ has shortcomings when viewed as a utility.

\subsection{A Supplementary Utility}

The criterion $p\geq p^*$ should be kept as the specification of risk tolerance. To overcome the 
objections to $p$ as a utility, we add to our optimization constraints a supplementary utility tolerance 
which must be met as well.  We define $U$ as the expectation of the following 
utility function,
\begin{equation}
\label{eqn6}
U(Y)=1-e^{-(Y-Y_M)/\Delta Y_M},\;U=\langle U(Y)\rangle.
\end{equation}
In Eq. (\ref{eqn6}), $\Delta Y_M$, introduced in section 4.1, is a natural
choice for \it utility sensitivity\rm.  In contrast to the Heaviside function, this function, 
while not unique, has the required utility characteristics.
It penalizes failure to meet the goal by going negative and rapidly 
increasing in magnitude with decreasing yield below $Y_M$.  It has positive and 
diminishing marginal utility as well.
\section{Portfolio Optimization, Normally Distributed Yields}
For short enough horizons, $T_H$, the random yield, $Y$, of a portfolio specified by $\{X_i\}$ is still 
linearly related to $\{Y_i\}$, as in section 4.2.
Suppose now, that the $Y_i$ are correlated Gaussian random variables.  Given its linearity 
in the $Y_i$, $Y$ is then normally distributed with probability distribution
\begin{equation}
\label{eqn8}
P(Y)=e^{-\frac{(Y-\bar{Y})^2} {2 V^2}}/\sqrt{2\pi V^2}.
\end{equation}
\subsection{The Risk Boundary}

The definition in Eq. (\ref{eqn4}) of $p$ now becomes
\begin{equation}
\label{eqn9}
p=\int^\infty_{Y_M} dY\:\frac{ e^{-(Y-\bar{Y})^2 / 2 V^2}}  {\sqrt{2 \pi V^2}}.
\end{equation}
Introducing $z$, a modified Sharpe's ratio\cite{sharpe},
\begin{equation}
\label{eqn10}
z=(\bar{Y}-Y_M)/ \sqrt{2} V,
\end{equation}
allows us to rewrite Eq. (\ref{eqn9}) as
\begin{equation}
\label{eqn11}
p=(1+erf(z))/2.
\end{equation}
The minimum acceptable value $p^*$ of $p$ thus defines through Eq. (\ref{eqn11}) a minimum 
acceptable value of $z$, $z^*$.  For example, if $p^*=0.9$ then $z^*=0.906$. Specifying that $p=p^*$ 
introduces a linear risk boundary
\begin{equation}
\label{eqn12}
\bar{Y}=Y_M +\sqrt{2}z^*V
\end{equation}
in the $\bar{Y}-V$ plane, as shown in Fig. (1).  Portfolios having acceptable risk, $p\geq p^*$, lie above 
the boundary.  Those with unacceptable risk, $p<p^*$, lie below it.

\subsection{The Utility Boundary}

From Eq. (\ref{eqn6}) and Eq. (\ref{eqn8}) the expected utility, $U$, is simply
\begin{equation}
\label{eqn13}
U=1-e^{-w},
\end{equation}
where the \it utility ratio \rm $w$ is
\begin{equation}
\label{eqn14}
w=(\bar{Y}-Y_M-{V^2}/{2\Delta Y_M})/\Delta Y_M.
\end{equation}
A minimum acceptable value $U^*$ of $U$ implies a minimum 
acceptable value $w^*$ of $w$.  For example, $U^*=0$ implies $w^*=0$.  Specifying $U^*$ 
defines the utility boundary
\begin{equation}
\label{eqn15}
\bar{Y}=Y_M+w^* \Delta Y_M+{V^2}/{2 \Delta Y_M}
\end{equation}
in the $\bar{Y}-V$ plane, also shown in Fig. 1.  Above the boundary, portfolios have $U\geq U^*$ and are 
acceptable.  Below, they have $U<U^*$ and are unacceptable. 

\subsection{The Acceptability Boundary}
A portfolio is fully acceptable if it meets both the risk and utility
criteria, that is if $p\geq p^*\;(z\geq z^*)$ and $U\geq U^*\;(w\geq w^*$). Depending on the value of
$w^*$ for a given $z^*$, the resulting acceptability boundary in the $\bar{Y}-V$ plane can coincide
with the utility boundary or can have one risk boundary and one or two utility boundary segments as illustrated in Fig. 1. The aceptability boundary is always effectively convex.
\subsection{Portfolio Optimization}

Portfolio optimization can now proceed by finding 
\begin{equation}
\label{eqn16}
sup_{\{X_i\}} \bar{Y}=\sum_i X_i \bar{Y}_i \;\;s.t.\;\;\sum_i X_i=1,\;p\ge p^*,\;U\ge U^*.
\end{equation}
The acceptability boundary is convex, and the efficient frontier is concave.  Consequently, 
there are zero, one, or two intersections as illustrated in Fig. 1.  If there are no 
intersections because the acceptability boundary lies above the efficient frontier, no 
portfolio can be constructed which meets the investment criteria.  If there are two 
intersections, the upper intersection specifies the optimal portfolio.  Because of the 
convex/concave characteristics of the boundaries and the monotonic upward shift of the 
acceptability boundary with increasing $p$ and $U$, this optimal portfolio has maximum 
allowable risk and minimum allowable utility.  One intersection is the marginal case.  
\begin{figure}
\includegraphics[width=15cm]{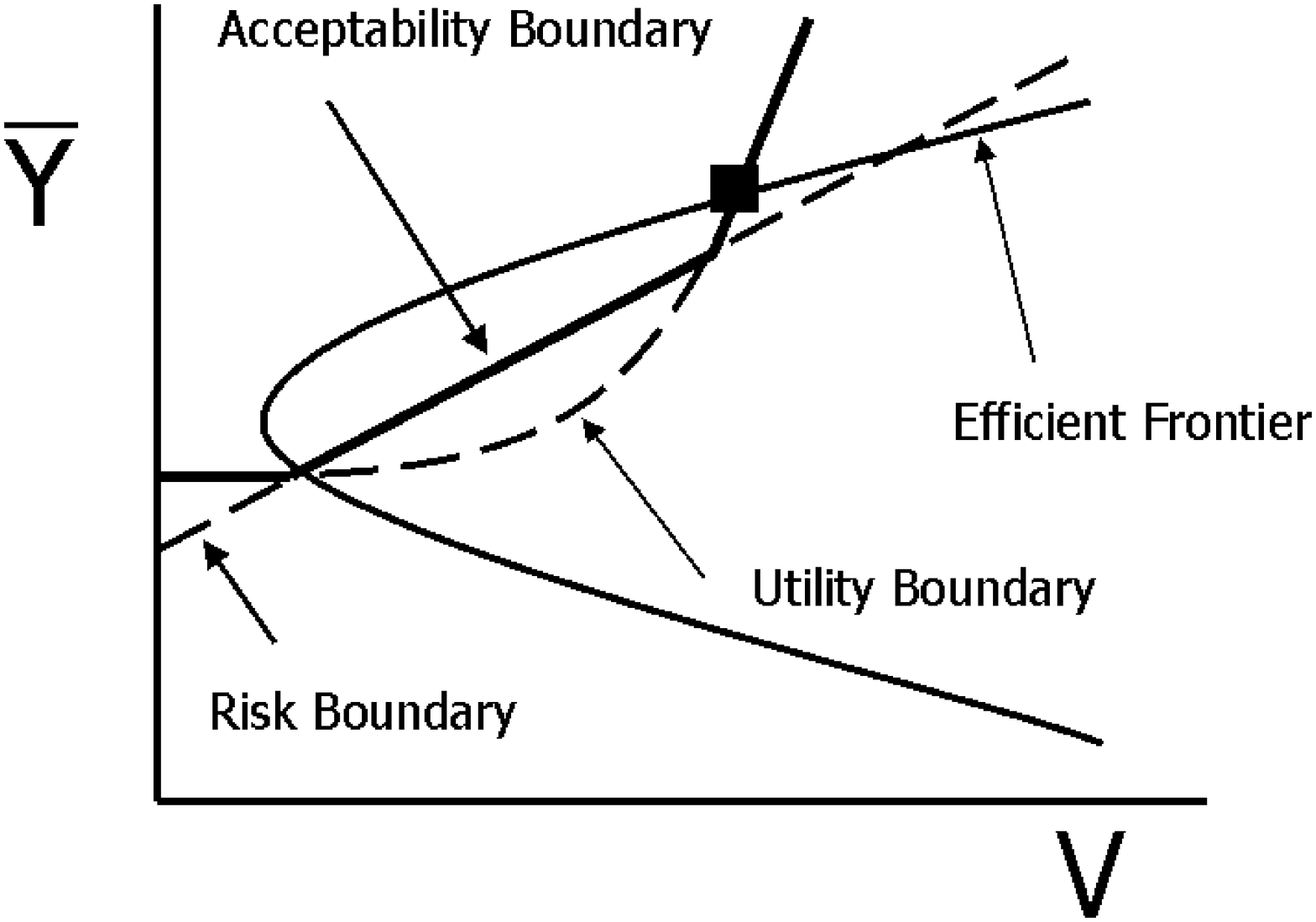}
\label{figure1}
\caption{The efficient frontier, the risk boundary, the
utility boundary line and the acceptability boundary
in the $\bar{Y}$-V plane.  The optimal portfolio is indicated by a solid square at the intersection of the acceptability boundary and the efficient frontier.}
\end{figure}

\section{The $p-U$ Plane}

\subsection{Setting The Investment Criteria}

We choose $Y^0_M$ to be 6\% and $\Delta Y_M$ to be 5\% so that $Y_M$ is 11\%.  We explore holding periods 
between 1 and 5 years and set a conservative value of $p^*$=0.9 and 
an aggressive value of $U^*$=0.  We do not allow short positions, $0\le X_i\le 1$.
We select stocks as the candidate assets.  Our historical data 
source is the daily closing prices of 13,000 NYSE and NSADAQ stocks during the 20 
year period 1977-1996 obtained from Genesis Financial Services.

\subsection{Scatter Plot Results}

The risk and utility criteria divide the $p-U$ plane into four sectors:

\begin{enumerate}
\item[A)] $p<p^*$, $U<U^*$-unacceptable $p$ and $U$ values.
\item[B)] $p<p^*$, $U\geq U^*$-unacceptable $p$ and acceptable $U$ values.
\item[C)] $p\geq p^*$, $U<U^*$-acceptable $p$ and unacceptable $U$ values.
\item[D)] $p\geq p^*$, $U\geq U^*$-acceptable $p$ and $U$ values.
\end{enumerate}

Typical scatter-plot results are shown in Table I for $p^*=0.9$ and $U^*=0$.  The entries in 
each sector column give the number of stocks from among those in the entire universe of 
13,000.  $T_D$ is the time span of the working data.  The proportion of stocks which individually meet the $p^*$ and $U^*$ 
criteria is small.  The total population of sectors B and D, where $U\geq U^*$ is substantially larger 
than that of sectors C and D where, $p\geq p^*$, illustrating the more aggressive character of the utility 
criterion.

\begin{table}
\caption{Sector Populations}
\begin{tabular}{|c|c|cccc|}\cline{3-6}
\multicolumn{2}{c}{}		&\multicolumn{4}{|c|}{Sectors}\\ \hline
$T_D$(Yrs)	&$T_H$(Yrs)	&A	&B	&C	&D \\ \hline
20		&3.5		&250	&1	&0	&0 \\ \hline
15		&3.5		&1007	&10	&3	&0 \\ \hline
15		&5.0		&958	&42	&18	&2 \\ \hline
\end{tabular}
\end{table}

\section{Results}
Optimizing $p$ instead of $\bar{Y}$ as an alternative to section 6.4,
we now construct an illustrative example of portfolio optimization.  The investment 
criteria we use are $Y_M$=11\%, $\Delta Y$=5\%, $T_H=5$ years, $T_D$=10 years, $p^*$=0.9, 
$U^*=0$, $0\le X_i\le 1$,
and $\{A_i\}$ comprises a set of 20 stocks drawn from the 129 
stocks in sector D, all of which have $p_i\geq p^*$ and $U_i\geq U^*$.  
We now find $\{X_i\}_{opt}$ as $arg\; sup_{\{X_i\}}\;p\;\;s.t.\;\;0\le X_i\le1,\;\sum_i X_i=1$
from which we evaluate $p_{opt}$, ${U}_{opt}$, and $\bar{Y}_{opt}$. The results are $p_{opt}=1.0$, ${U}_{opt}=.918$, and $\bar{Y}_{opt}\ge 25.0\%$ and $V_{opt}=4.5\%$. 
Only 6 of the 20 stocks have nonzero allocation ratios:  
$X_3$=0.35, $X_6$=0.043, $X_{12}$=0.043, $X_{13}$=0.185, $X_{17}$=0.336, and $X_{18}$=0.043. 

\section{Summary and Conclusions}

We started with a brief summary of MPT to introduce the concepts of the efficient frontier, the 
capital market line, and the market portfolio.  We then argued that the concept of a risk-free return is invalid for longer holding periods.
To replace the volatility, 
which measures uncertainty not risk, we introduced the probability of failure to meet a 
preset investment goal as a measure of risk.  The corresponding probability of success, $p$, 
is a utility which neither penalizes failure nor incorporates diminishing positive marginal 
utility.  We supplement $p$ with an appropriately defined utility $U$ and impose minimum 
acceptable values of $p$ and $U$ for the portfolio.  To explore the feasibility of 
implementing $p-U$ based portfolio optimization, we computed the $p_i$ and $U_i$ values for 
individual stocks over various holding periods using 
historical data drawn from a database of 13,000 stocks.  Composing the asset set of 20 stocks 
from the acceptable sector of the $p-U$ plane, we optimized the 
probability of success for a lower bound to the expected yield.  The results imply the feasibility of constructing a convex $\bar{Y}-p$ efficient frontier 
in the $\bar{Y}-p$ plane.  The optimal portfolio can be that which maximizes $\bar{Y}$ on the frontier subject 
to $p\geq p^*$ and $U\geq U^*$ or simply that which maximizes $p$.  The $U_i$, $p_i$ scatter plot is a powerful tool for candidate asset selection. 

All of this is an academic exercise unless it is accompanied by a measure of confidence 
that the use of historical data generates predictive power.  Fundamental analysis of the 
candidate companies, industries, etc., must therefore be an essential component of 
portfolio construction.






\end{document}